\documentclass[11pt]{article}

\usepackage[margin=1in]{geometry}
\usepackage{amsmath,amssymb}
\usepackage{booktabs}
\usepackage{array}
\usepackage{graphicx}
\usepackage{hyperref}
\usepackage{cleveref}
\usepackage{authblk}

\title{Validation of a high-order finite difference compressible solver}
\author[1]{Yujoo Kang}
\author[1,*]{Sang Lee}
\affil[1]{Aerospace Engineering, Korea Advanced Institute of Science and Technology, Daejeon, 34141, South Korea}
\affil[*]{Corresponding author: Sang Lee, slee1@kaist.ac.kr}

\date{}

\begin{document}
\maketitle

\begin{abstract}
The verification and validation of a high-order compressible in-house solver based on a compact finite difference scheme are presented. Validation is performed using five canonical cases: the one-dimensional Sod shock tube problem, two-dimensional shock--shear layer interaction, compressible channel flow, compressible turbulent boundary layer, and shock-turbulent boundary layer interaction. Comparisons against exact solutions and reference direct numerical simulation data demonstrate accurate shock capturing, resolution of vortical structures, and good agreement for first and second order statistics.
\end{abstract}

\section{Introduction}
The hyperbolic nature of the compressible Navier–Stokes equations gives rise to propagating discontinuities, such as shock waves, whose interaction with turbulence poses significant challenges for high-fidelity simulations. Turbulence-resolving methods require minimal numerical dissipation to preserve small-scale structures, whereas shock-capturing schemes rely on additional dissipation to stabilize discontinuities. We therefore develop a high-order finite difference compressible solver designed to accurately resolve small-scale turbulent structures while maintaining the robustness.

The compact finite-difference method \cite{Lele_1992_JCP} employs implicit relations between neighboring grid points to compute spatial derivatives, achieving high-order accuracy with spectral-like resolution using compact stencils. Because of superior dispersion and dissipation characteristics, compact schemes enable accurate resolution of small-scale turbulent structures. However, the implicit coupling of spatial derivatives in compact schemes requires the solution of tridiagonal systems, typically handled using the Thomas algorithm, which increases the computational cost compared to explicit central schemes. To address this limitation, an efficient parallel algorithm based on the Message Passing Interface is implemented. Through domain decomposition and parallel tridiagonal solvers, the computational overhead of compact discretizations is effectively mitigated, enabling scalable large-scale simulations on both CPU- and GPU-based HPC clusters \cite{Kim_2021_PASCAL,Yang_2023_JPC, Kim_2025_arXiv}.

Here, we present a high-order compressible finite-difference solver based on compact finite-difference schemes, capable of simulating compressible flows over relatively complex geometries, such as curved or faceted walls. The solver is verified and validated through five benchmark tests. 

\section{Code descriptions}
\label{sec:method}
The governing equations for the in-house codes \cite{Kang_2024_PoF} are the three-dimensional, unsteady, compressible Navier-Stokes equations, expressed as follows:
\begin{align}
\frac{\partial \rho}{\partial t} + \frac{\partial}{\partial x_j} (\rho u_j) &= 0\ , \\
\frac{\partial}{\partial t}(\rho u_i) + \frac{\partial}{\partial x_j}(\rho u_j u_i + p\delta_{ij}-\tau_{ij}) &= f_i\ , \\
\frac{\partial E}{\partial t} + \frac{\partial}{\partial x_j}(u_j E - S_{ij} u_i + q_j) &= f_i u_i\ .
\end{align}
The coordinate system is defined as $x_i = (x, y, z)$, corresponding to the streamwise, wall-normal, and spanwise directions, respectively. $\rho$ denotes the density, $u_i=(u,v,w)$ is the velocity corresponding to the $x_i=(x,y,z)$ direction, $p$ is the pressure, $E = \rho (e + u_i u_i/2)$ presents the total fluid energy in which $e$ is the internal energy, $S_{ij}$ presents the strain rate tensor and $q_i$ denotes the heat flux. $f_i$ is the external force. The viscous stress tensor, $\tau_{ij}$, is determined by the Newtonian fluid relationship:
\begin{equation}
\tau_{ij} = \mu \left( \frac{\partial u_i}{\partial x_j} + \frac{\partial u_j}{\partial x_i} \right)  - \frac{2}{3} \mu \delta_{ij} \frac{\partial u_k}{\partial x_k}\ ,
\end{equation}
where $\mu$ is the dynamic viscosity. The viscosity follows Sutherland's law, while the thermal conductivity is computed using the Prandtl relation, $\kappa=\mu c_p /Pr$, where $Pr$ is the Prandtl number and $c_p$ is the heat capacity at constant pressure.

The governing equations are solved in conservative form using generalized coordinates $(\xi, \eta, \zeta)$, which are transformed from Cartesian coordinates $(x, y, z)$:
\begin{equation}
\frac{\partial \bar{U}}{\partial t}+\frac{\partial \bar{E}}{\partial \xi}+\frac{\partial \bar{F}}{\partial \eta} + \frac{\partial \bar{G}}{\partial \zeta}={\bar{S}}_{ext} \ ,
\end{equation}
where $\bar{U}$ is the solution vector, $\bar{S}_{\text{ext}}$ is the external body force vector, and $\bar{E}$, $\bar{F}$, and $\bar{G}$ are the flux terms, including both convective and viscous contributions. These variables are defined as follows:
\begin{align}
U = 
\begin{bmatrix}
\rho \\
\rho u \\
\rho v \\
\rho w \\
\rho e
\end{bmatrix}, \quad
S_{ext} =
\begin{bmatrix}
0 \\
f_1 \\
f_2 \\
f_3 \\
f_1 u + f_2 v + f_3 w
\end{bmatrix}, \quad
E =
\begin{bmatrix}
\rho u \\
\rho u^2 + p - \tau_{11} \\
\rho u v - \tau_{12} \\
\rho u w - \tau_{13} \\
(\rho e + p)u - (u \tau_{11} + v \tau_{12} + w \tau_{13}) + q_1
\end{bmatrix}.
\end{align}
The bar denotes the converted vector in the generalized coordinate: $\bar{U}=U/J$, 
 $\bar{S}_{ext}=S_{ext}/J$, and the converted flux terms are formulated following:
\begin{equation}
\bar{E}=\frac{1}{J} \left(\frac{\partial \xi}{\partial x}E + \frac{\partial \xi}{\partial y}F + \frac{\partial \xi}{\partial z}G\right) \ ,
\end{equation}
where $J$ represents the Jacobian of the transformation $J=| \partial (x,y,z)/\partial (\xi,\eta,\zeta)|$.

Spatial derivatives are computed using a sixth-order compact finite-difference scheme developed by Lele \cite{Lele_1992_JCP}, which provides spectral-like resolution. Time integration is performed using a third-order explicit Runge–Kutta method, together with high-order Padé-type non-dispersive spatial filters \cite{Gaitonde_2000_AIAA, Visbal_2002_JCP}, which are applied to the solution vector at the final Runge–Kutta stage. To mitigate numerical oscillations, a localized artificial diffusivity method developed by Kawai and Lele \cite{Lele_2008_JCP} is employed.

\section{Results}
\label{sec:benchmark}
Five benchmark cases corresponding to canonical compressible-flow validation cases are carried out: a one-dimensional Sod shock tube, a two-dimensional shock--shear layer interaction, compressible turbulent channel flow, a compressible turbulent boundary layer, and a shock-turbulent boundary layer interaction. The fluid is assumed to be a perfect gas with a heat capacity ratio of 1.4 and $Pr=0.72$. Throughout this section, the superscript $^{*}$ represents non-dimensionalized parameters, while subscription $_{\infty}$ and $_{w}$ denote the inflow and wall property, respectively.

\subsection{One-dimensional Sod shock-tube problem}
\label{sec:case1}

The classical one-dimensional Sod shock-tube problem\cite{Sod_1978_JCP} is conducted to assess the shock-capturing capability of the numerical schemes. The equations are solved with standard Riemann initial conditions consisting of a left-moving rarefaction wave, a contact discontinuity, and a right-moving shock. The initial conditions are:
\begin{eqnarray}
(\rho^*, u^*, v^*, p^*)^T
&=&
\begin{cases}
(1.0,\; 0.0,\; 1.0,\; 1.4)^T, & x \in [-0.5,\,0], \\
(0.125,\; 0.0,\; 0.1,\; 1.4)^T, & x \in [0,\,0.5].
\end{cases}
\end{eqnarray}
Simulations are performed using  200, 400, and 600 uniformly spaced grid points. The numerical results are compared with the exact Riemann solution and with classical fifth-order weighted essentially non-oscillatory (WENO) results\cite{Jiang_1996_JCP}.

\begin{figure*}[htb!]
    \centering
    \includegraphics[width=0.6\linewidth]{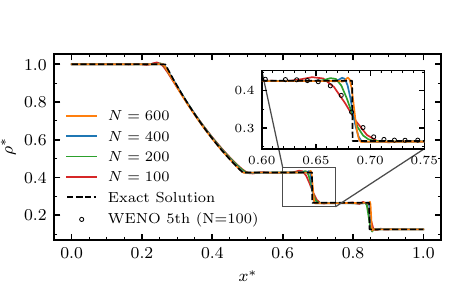}
    \caption{\label{fig:sod} Density profiles for the one-dimensional Sod shock-tube problem at $t^*=0.2$, comparing numerical solutions obtained with different grid resolutions (N=100, 200, 400, and 600  ) against the exact Riemann solution and 5th order WENO results by Jiang and Shu\cite{Jiang_1996_JCP}.}
\end{figure*}

Figure \ref{fig:sod} shows the density distribution at $t^*=0.2$, comparing numerical solutions at different grid resolutions with the exact Riemann solution and fifth-order WENO results\cite{Jiang_1996_JCP}. All simulations correctly capture the main wave structures, including the shock wave, contact discontinuity, and expansion fan. As the grid resolution increases, the numerical solution exhibits clear convergence toward the exact solution. The absence of significant spurious oscillations near discontinuities demonstrates the robustness of the shock-capturing and its ability to accurately resolve the strong gradients while preserving the shock strength.

\subsection{Two-dimensional shock--shear layer interaction}
\label{sec:case2}

The two-dimensional problem of shock-wave impingement on a spatially evolving mixing layer is considered to examine the resolution of small-scale vortical structures interacting with a shock discontinuity\cite{Sandham_1998}. The computational domain spans $[0,200] \times [-20,20]$, discretized with $500 \times 100$ grid points. At the inlet, the streamwise velocity profile of the mixing layer is prescribed as
\begin{equation}
u^*(y^*) = 2.5 + 0.5 \tanh(2y^*).
\end{equation}
The pressure is initialized uniformly as $p^* = 0.3327$. The density is specified as $\rho^* = 1.6374$ for $y^* > 0$ and $\rho^* = 0.3626$ for $y^* < 0$. To promote the development of the mixing layer, a time-dependent transverse velocity perturbation is imposed at the inlet:
\begin{equation}
v^{*\prime}(y,t) =
\sum_{k=1}^{2}
a_k
\cos\left( 2\pi k \frac{t^*}{T} + \phi_k \right)
\exp\left(-\frac{y^{*2}}{b}\right),
\end{equation}
where $a_1 = a_2 = 0.05$, $\phi_1 = 0$, $\phi_2 = \pi/2$, and $b = 10$. The forcing period is defined as $T = \lambda/u_c$, with $\lambda = 30$ and $u_c = 2.68$. An oblique shock is introduced by prescribing post-shock conditions $(\rho^*,u^*,v^*,p^*) = (2.1101,\,2.9709,\,-0.1367,\,0.4754)$
at the upper boundary. The lower boundary is treated as a slip wall. The Reynolds number is $Re = 500$, and the Mach numbers of the upper- and lower-stream inflows are 5.6250 and 1.7647, respectively.

\begin{figure*}[htb!]
    \centering
    \includegraphics[width=1.0\linewidth]{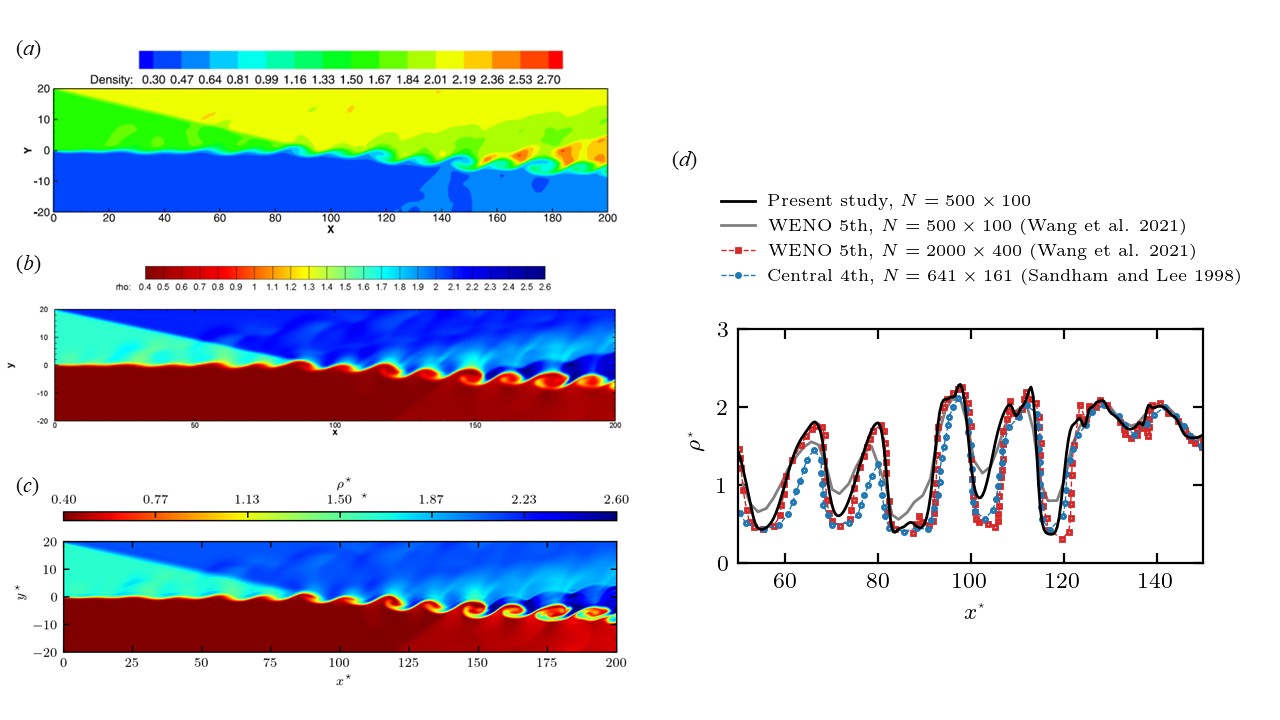}
    \caption{\label{fig:2Dshock} Instantaneous density contour at $t^*=120$ by (a) 5th order WENO scheme by Wang \textit{et al.}\cite{Wang_2021_CMAME}, (b) 5th order WENO-Z scheme by Peng \textit{et al.}\cite{Peng_2019_CaF} and (c) Present study based on 6th order compact scheme. (d) Density distribution at $y^*=0$.}
\end{figure*}

Figure \ref{fig:2Dshock} (c) shows the density distribution at $t^* = 120$, compared with the results obtained using a fifth-order WENO scheme by Wang \textit{et al.}\cite{Wang_2021_CMAME} (Figure \ref{fig:2Dshock}(a)) and a fifth-order WENO-Z scheme by Peng\textit{et al.} \cite{Peng_2019_CaF}(Figure \ref{fig:2Dshock} (b)) on the same grid. Overall, the density distribution is consistent with benchmark results reported in previous studies. In particular, the vortical structures generated after the shock interaction are more clearly captured than those obtained using the WENO-based schemes. Figure \ref{fig:2Dshock}(d) shows the density distribution along the centerline $y^* = 0$. The present results are compared with those obtained using a fifth-order WENO scheme on the same and finer grids by Wang \textit{et al.}\cite{Wang_2021_CMAME}, and a fourth-order central-difference scheme on a finer grid by Sandham and Lee\cite{Sandham_1998}. The density distribution generally falls within the range of the other simulation results. When compared with the fifth-order WENO scheme on the same grid, the density fluctuations associated with the vortical structures are predicted at a level closer to those obtained on the finer grid.

\subsection{Compressible channel flow}
\label{sec:case3}

A direct numerical simulation (DNS) of compressible channel flow is conducted at a bulk Mach number of $M_b = 1.5$ and a bulk Reynolds number of $Re_b = 6000$, corresponding to a friction Reynolds number of $Re_\tau \approx 218$. These flow conditions are consistent with the simulations reported by Morrison \textit{et al.}\cite{Morinishi_2004_JFM} and Modesti and Pirozzoli\cite{Modesti_2016_IJHF}.  The computational domain is a rectangular box with dimensions $(L^*_x \times L^*_y \times L^*_z) = (4\pi \times 2 \times 4\pi/3)$. The domain is discretized using $N_x \times N_y \times N_z = (120 \times 180 \times 120)$ for the base case, and $N_x \times N_y \times N_z = (256 \times 200 \times 230)$ for the finer case. The simulation settings for the present study and other DNS studies\cite{Morinishi_2004_JFM,Modesti_2016_IJHF} are listed in Table \ref{tab:channel_dns}

The initial condition consists of a parabolic streamwise velocity profile superposed with random perturbations in $u, v$, and $w$, together with large-scale sinusoidal perturbations in $v$ and $w$,  corresponding to a streamwise-aligned roller. A forcing term $f_i$ is evaluated at each time step to discretely enforce a constant mass-flow rate in time, and the associated term is added to the right-hand side of the energy equation. The wall-normal boundaries are treated as isothermal walls, while the streamwise and spanwise boundaries are periodic.

\begin{table}[t]
\centering
\caption{Flow parameters and grid information of previous studies and the present simulation.}
\label{tab:channel_dns}
\begin{tabular}{lccccccc}
\toprule
Case & $M_b$ & $Re_\tau$ & $N_x$ & $N_y$ & $N_z$ & $\Delta x^+$ & $\Delta z^+$ \\
\midrule
Modesti and Pirozzoli (2016)\cite{Modesti_2016_IJHF} & 1.5 & 218 & 120 & 180 & 120 & 23 & 7.6 \\
Morrison \textit{et al.}~(2004)\cite{Morinishi_2004_JFM}      & 1.5 & 218 & 120 & 180 & 120 & 23 & 7.6 \\
Present study (base)       & 1.5 & 218 & 120 & 180 & 120 & 23 & 7.6 \\
Present study (fine)       & 1.5 & 218 & 256 & 200 & 230 & 10 & 3.9 \\
\bottomrule
\end{tabular}
\end{table}

\begin{figure*}[htb!]
    \centering
    \includegraphics[width=0.5\linewidth]{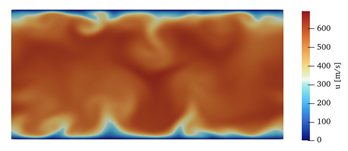}
    \caption{\label{fig:channel_paraview} Instantaneous streamwise velocity contour at the cross section. }
\end{figure*}

\begin{figure*}[htb!]
    \centering
    \includegraphics[width=1.0\linewidth]{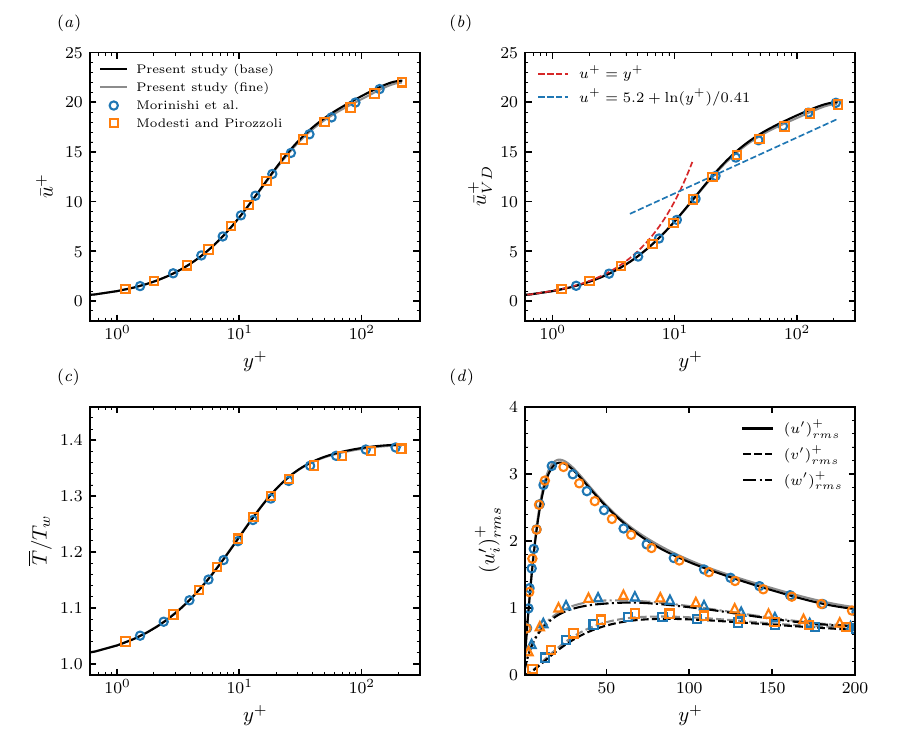}
    \caption{\label{fig:channel} Comparison of compressible channel flow with other DNS data \cite{Morinishi_2004_JFM, Modesti_2016_IJHF}: (a) Mean velocity, (b) Van-Driest transformed mean velocity, (c) normalized mean temperature and (d) root mean square (rms) velocity fluctuations. }
\end{figure*}

Figure \ref{fig:channel_paraview} shows the instantaneous velocity contour of the channel flow, and Figure \ref{fig:channel} presents comparisons of the present DNS results with reference data\cite{Morinishi_2004_JFM,Modesti_2016_IJHF}. The profiles are averaged in time and spanwise direction, denoted by ( $\bar{ }$ ). The results obtained on the base and finer grids are in excellent agreement, indicating that the solution is grid-converged for the quantities of interest. Figures \ref{fig:channel}(a) and (b) show the mean streamwise velocity profiles expressed in inner scaling using wall units and the Van-Driest transformation, respectively. The agreement with the reference data is shown throughout the viscous sublayer, buffer layer, and logarithmic region, indicating that the near-wall resolution and compressibility treatment are adequate. Figure \ref{fig:channel}(c) compares the mean temperature profile normalized by the wall temperature, showing a good agreement across the wall-normal extent. Figure \ref{fig:channel}(d) shows the root-mean-square velocity fluctuations in wall units. The peak locations and magnitudes of the streamwise, wall-normal, and spanwise components are well reproduced.

\subsection{Compressible turbulent boundary layer}
\label{sec:case4}

The compressible turbulent boundary layer with freestream Mach number of $M_\infty = 2.25$ and Reynolds number based on momentum thickness of $Re_\theta \approx 2100$ is simulated using a turbulent tripping method. The domain length is $(L_x \times L_y \times L_z) = (75,\,5,\,7.5)\delta_i$, and a fully developed turbulent boundary layer is obtained at $x/\delta_i \approx 70$, where $\delta_i$ is the boundary layer thickness at the inlet. The grid resolution is $N_x \times N_y \times N_z = (1300 \times 250 \times 301)$,
corresponding to grid spacings of $(\Delta x^+,\, \Delta y_w^+,\, \Delta z^+) = (16,\,0.5,\,6)$.
The turbulent boundary layer is generated based on the configuration of Poggie \textit{et al.}\cite{Poggie_2015_CaF}, who employed a turbulent tripping method with a domain length $(L_x \times L_y \times L_z) = (100,\,5,\,50)\delta_i$ and obtained a reference turbulent boundary layer at $x/\delta_i \approx 100$.

\begin{figure*}[htb!]
    \centering
    \includegraphics[width=1.0\linewidth]{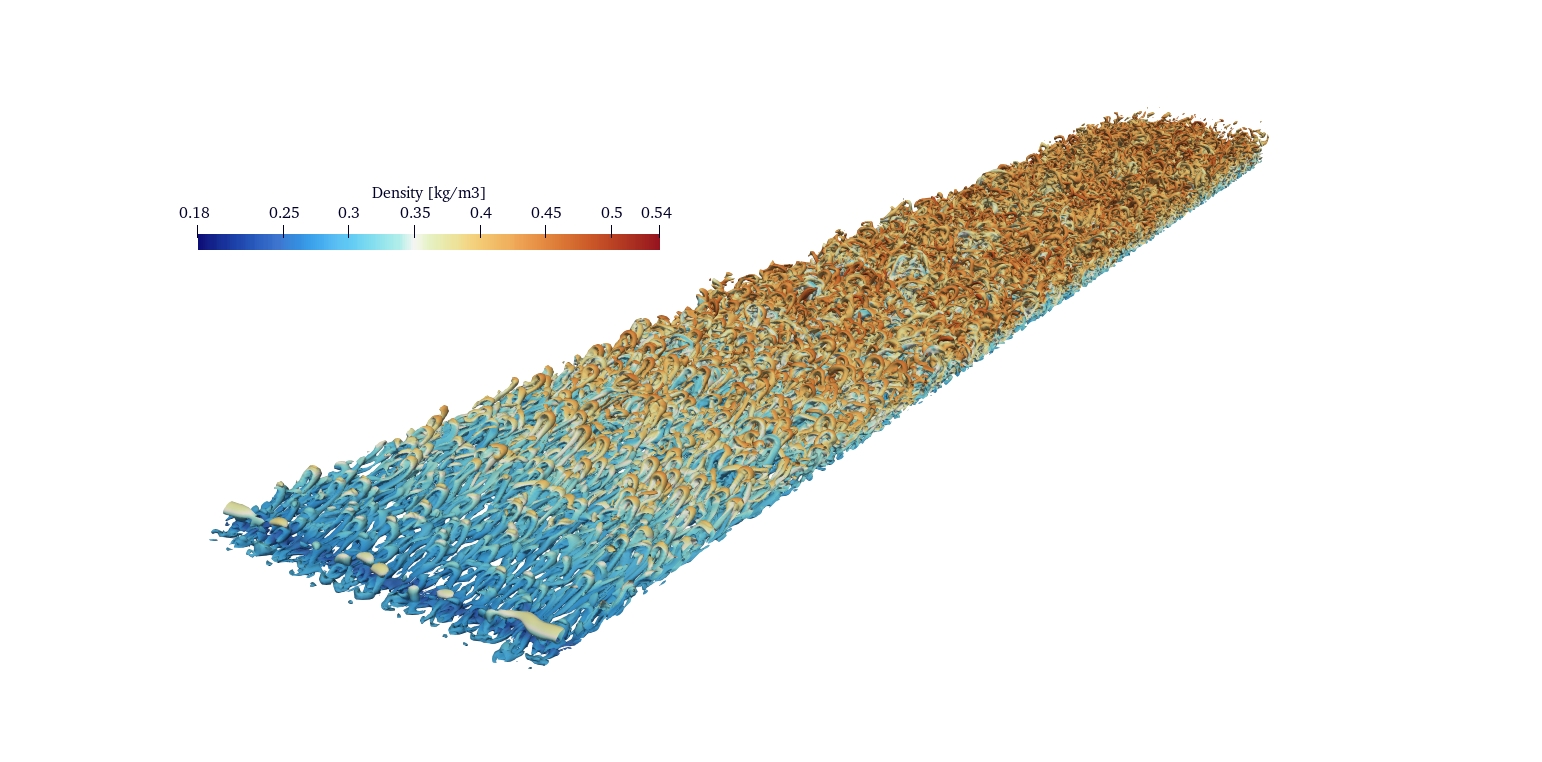}
    \caption{\label{fig:bl_q} Iso-surface of Q criterion colored by density of current simulation. }
\end{figure*}

The wall is treated as a non-slip adiabatic boundary. Periodic boundary conditions are applied in the spanwise direction, the outlet flow is extrapolated from the interior, and a Blasius boundary-layer profile is prescribed at the inlet. The transition to turbulent flow is achieved using the counter-flow body-force tripping method by Poggie and Smiths\cite{Poggie_2001_JFM}, as shown in Figure \ref{fig:bl_q}. The body force $f$ is given: 
\begin{equation}
f = \frac{2D_c}{\pi l_1 l_2 l_3} \sin^2\left(\pi \frac{z-X_3}{l_3}\right) 
\exp\left[-\left(\frac{x-X_1}{l_1}\right)^2 - \left(\frac{y-X_2}{l_2}\right)^2\right].
\end{equation}
The force distribution exhibits a Gaussian profile in the streamwise and wall-normal direction, while in the spanwise direction, it follows the square of the sine function. Here, ($X_1$, $X_2$, $X_3$)=(2.5,0,0)$\delta_i$ denote the center coordinates of the forcing region, and $(l_1,l_2,l_3)=(0.17,0.01,0.5)\delta_i$ represent the spatial extent of the forcing region, both of which are user-defined parameter. The parameter $D_c$ controls the magnitude of the body force. The body force is decomposed as $f_i=(f\cos179^\circ, f\sin179^\circ,0)$, ensuring that the force direction is nearly parallel to the wall. 

\begin{table}[t]
\centering
\caption{Flow parameters and grid information of previous DNS studies and the present simulation.}
\label{tab:turbulent_bl_dns}
\begin{tabular}{lccccccc}
\toprule
Case & $M_\infty$ & $Re_\theta$ & $N_x$ & $N_y$ & $N_z$ & $\Delta x^+$ & $\Delta z^+$ \\
\midrule
Pirozzoli \textit{\textit{et al.}}~(2011)\cite{Pirozzoli_2011_JFM}      & 2.0  & 2400 & 4160 & 221 & 400 & 6  & 5 \\
Poggie \textit{et al.}~(2015), fine\cite{Poggie_2015_CaF}  & 2.25 & 2000 & 11287 & 1277 & 568 & 2  & 2 \\
Poggie \textit{et al.}~(2015), coarse\cite{Poggie_2015_CaF}& 2.25 & 2000 & 2278 & 1277 & 117 & 10 & 10 \\
Schlatter and Örlü~(2010)\cite{Schlatter_2010_JFM}   & --   & 4300 & 8192 & 513 & 768 & 9  & 4 \\
Present study               & 2.25 & 2100 & 1300 & 250 & 301 & 16 & 6 \\
\bottomrule
\end{tabular}
\end{table}

\begin{figure*}[htb!]
    \centering
    \includegraphics[width=1.0\linewidth]{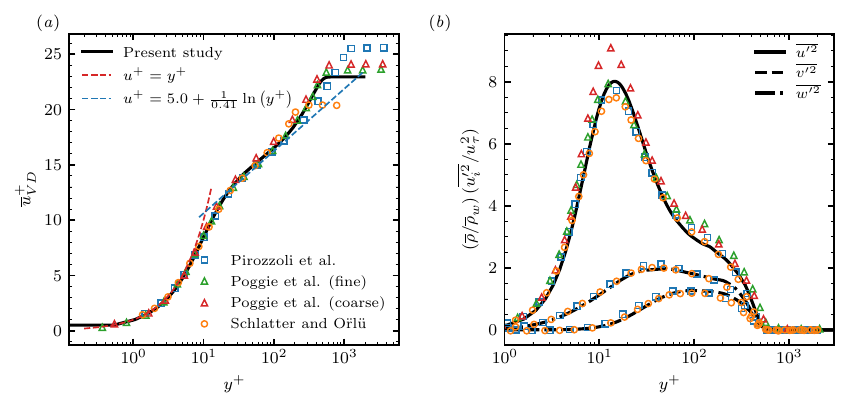}
    \caption{\label{fig:bl_prob} Comparison of compressible boundary layer results with other DNS data \cite{ Poggie_2015_CaF, Pirozzoli_2011_JFM,Schlatter_2010_JFM}: (a) Van-Driest transformed mean velocity and (b) density-scaled root mean square velocity fluctuations. }
\end{figure*}

The simulation settings of the reference DNS datasets are summarized in Table \ref{tab:turbulent_bl_dns}. For comparison, coarse- and fine-grid data from Poggie\textit{et al.}\cite{Poggie_2015_CaF} are compared with the present simulation results. In addition, compressible boundary-layer DNS data with $M_\infty = 2.0$ and $Re_\theta \approx 2400$ by Pirozzoli \textit{et al.}\cite{Pirozzoli_2011_JFM}, and incompressible boundary-layer DNS data with $Re_\theta \approx 4300$ by Schlatter and Örlü\cite{Schlatter_2010_JFM} are also compared with the present results.

Figure \ref{fig:bl_prob} shows the Van-Driest-transformed mean velocity and density-scaled Reynolds normal stress profiles compared with other compressible and incompressible DNS data. Overall, both the velocity and Reynolds normal stress profiles align well with the reference data. In particular, despite the relatively coarse grid resolution in the streamwise and spanwise directions, the peak magnitude of the streamwise Reynolds normal stress is well captured and closely matches the results from higher-resolution DNS studies.

\subsection{Shock-turbulent boundary layer interaction}
\begin{figure*}[htb!]
    \centering
    \includegraphics[width=1.0\linewidth]{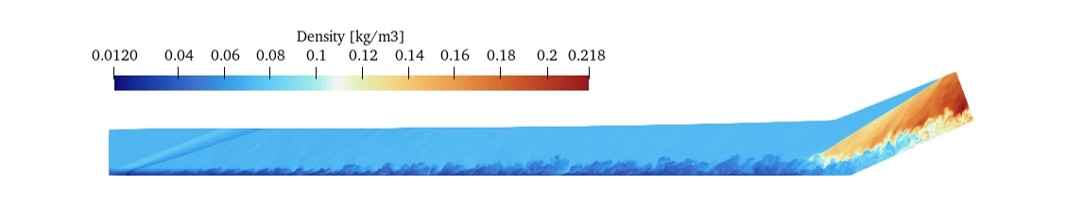}
    \caption{\label{fig:contour_STBLI} Instantaneous contour of density. }
\end{figure*}
A DNS of shock–turbulent boundary-layer interaction over a $24^\circ$ compression ramp is performed at $M_\infty \approx 2.9$ and $Re_\theta \approx 2400$, consistent with the experimental conditions of Bookey \textit{et al.}\cite{Bookey_2005_AIAA}. The flat plate has dimensions of $L_x \times L_z = 237.5 \times 35$ mm$^2$ ($\approx 35.4\delta_0 \times 5.2\delta_0$), and a ramp of length 54.5 mm ($\approx 8.1\delta_0$) with the same width. The computational domain consists of a structured grid of $N_x \times N_y \times N_z =2865 \times 298 \times 420$ grid points. The grid resolution satisfies $(\Delta x^+, \Delta y_w^+, \Delta z^+)=(4.6, 0.42, 4.2)$, as measured at $x_0$, which is located $5.37\delta_0$ upstream of the ramp corner. The subscript $_{0}$ denotes the quantities evaluated at $x_0$.

No-slip isothermal wall conditions are applied, with periodicity in the spanwise direction. The outflow is handled by interior extrapolation, while a Blasius boundary-layer profile is prescribed at the inlet, and turbulence is initiated using a counter-flow body-force tripping method.

\begin{table}[t]
\centering
\caption{Flow parameters of the reference experiment and the present DNS.}
\label{tab:bookey_comparison}
\begin{tabular}{lccccccc}
\toprule
Case & $Ma_\infty$ & $T_\infty$ [K] & $T_w$ [K] & $\delta_0$ [mm] & $\theta_0$ [mm] & $Re_\theta$ & $C_f$ \\
\midrule
Bookey \textit{et al.}~\cite{Bookey_2005_AIAA} & 2.9 & 107.1 & 307 & 6.7 & 0.43 & 2400 & 0.00225 \\
Present DNS                                   & 2.9 & 107.0 & 307 & 6.7 & 0.43 & 2400 & 0.00227 \\
\bottomrule
\end{tabular}
\end{table}

\begin{figure*}[htb!]
    \centering
    \includegraphics[width=1.0\linewidth]{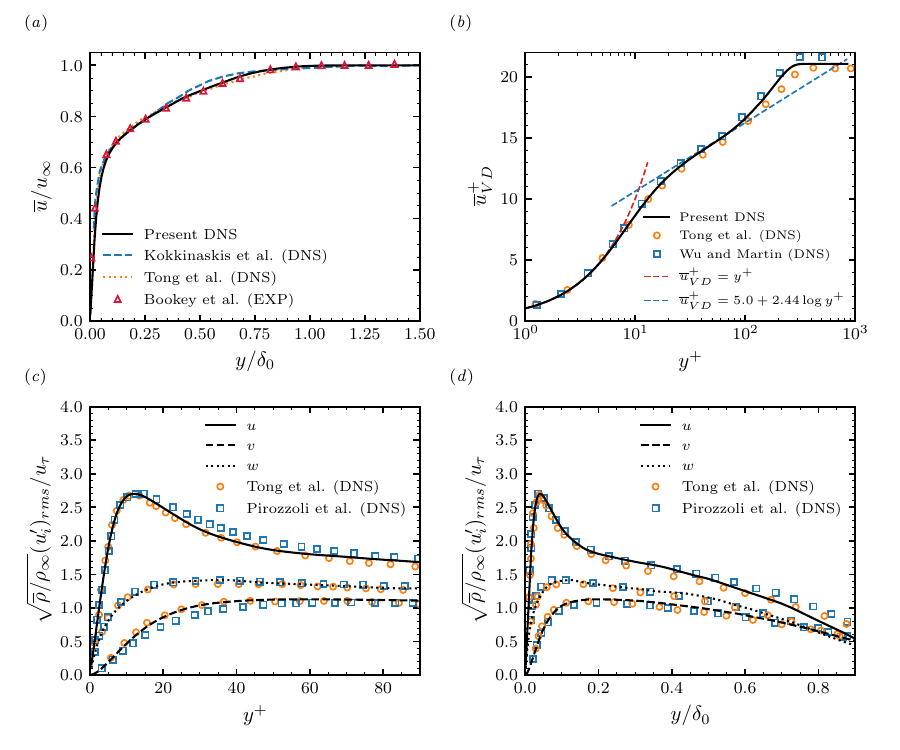}
    \caption{\label{fig:stbli_inflow} Comparison of (a) mean streamwise velocity profile, (b) Van-Driest transformed mean velocity, density-scaled turbulent fluctuation in (a) inner and (b) outer scale with experimental\cite{Bookey_2005_AIAA} and other DNS data\cite{Kokkinakis_2020_PoF,Tong_2017_PoF,Wu_2007_AIAA,Pirozzoli_2010_JFM} at $x_0$.}
\end{figure*}

\begin{figure*}[htb!]
    \centering
    \includegraphics[width=1.0\linewidth]{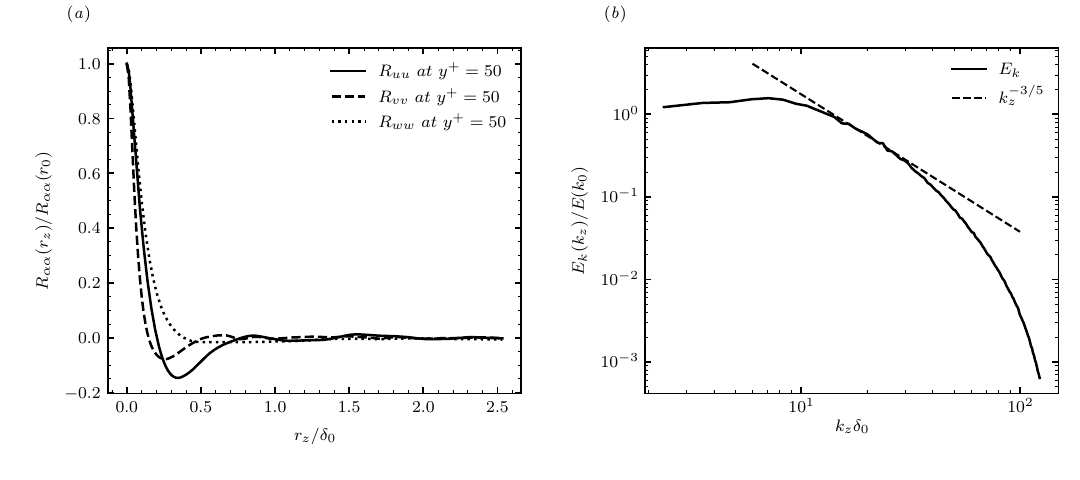}
    \caption{\label{fig:third} (a) Two-point spanwise autocorrelation and (b) One-dimensional energy spectra at $x_0$ }
\end{figure*}

\begin{figure*}[htb!]
    \centering
    \includegraphics[width=1.0\linewidth]{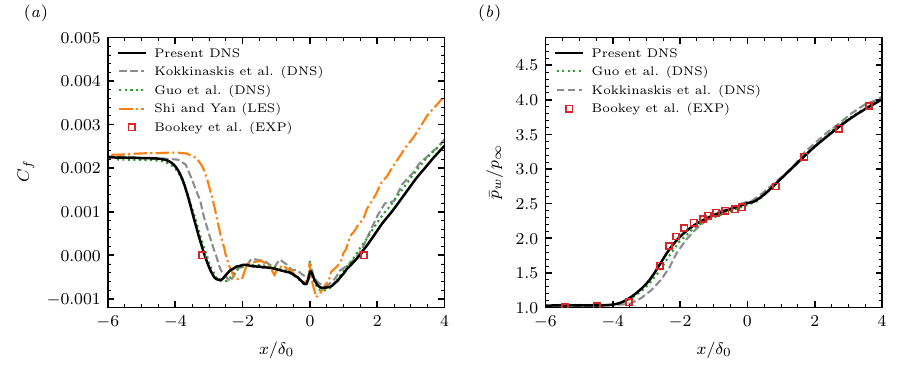}
    \caption{\label{fig:wall_property} Comparison of (a) skin friection coefficient and (b) wall pressure with experimental\cite{Bookey_2005_AIAA} and other DNS\cite{Kokkinakis_2020_PoF, Guo_2022_PoF} and LES\cite{Shi_2023_PoF} data.}
\end{figure*}

 The inflow turbulent statistics are obtained at $x_0$ and averaged in time and spanwise direction. The main inflow parameters are compared with the experimental data of Bookey et al.~\cite{Bookey_2005_AIAA} and are listed in Table~\ref{tab:bookey_comparison}. To ensure a consistent incoming flow, the inflow turbulent boundary layer is validated by comparing mean and fluctuation statistics with available experimental\cite{Bookey_2005_AIAA} and DNS data \cite{Kokkinakis_2020_PoF,Tong_2017_PoF,Wu_2007_AIAA,Pirozzoli_2010_JFM}, as shown as Figure~\ref{fig:stbli_inflow}. Overall, the inflow profiles agree well with the references, indicating that the upstream boundary layer is appropriately represented for subsequent interaction with the ramp shock.

Figure \ref{fig:third} (a) shows the two-point spanwise autocorrelation of three velocity components, $R_{\alpha\alpha}$ at $x_0$. All correlations decay to zero within approximately $0.5\delta_0$, indicating that the spanwise extent of the computational domain is sufficient to ensure statistical decorrelation of turbulent fluctuations. Figure~\ref{fig:third}(b) shows the normalized one-dimensional energy spectra measured at $y^+\approx50$. The absence of artificial spectral peaks confirms that no spurious length scales are introduced by the tripping procedure. The spectra exhibit the classical $k^{-5/3}$ inertial-range scaling, followed by a pronounced decay at high wavenumbers associated with turbulent dissipation. This behavior demonstrates that the adopted grid resolution is adequate to resolve the small-scale turbulent motions.

Figure~\ref{fig:wall_property} compares the skin-friction coefficient $C_f$ and the mean wall pressure obtained in the present study with available experimental~\cite{Bookey_2005_AIAA}, DNS~\cite{Kokkinakis_2020_PoF,Guo_2022_PoF}, and LES~\cite{Shi_2023_PoF} data. 
All reference datasets correspond to the same ramp configuration and flow conditions. A flow reversal region is clearly observed around the ramp corner, as indicated by the negative values of $C_f$. The separation and reattachment locations, defined by the zero-crossings of $C_f$, are in close agreement with the experimental measurements. In addition, the overall distributions of both $C_f$ and wall pressure show good agreement with previous numerical studies, with particularly close correspondence to the experimental \cite{Bookey_2005_AIAA} DNS results~\cite{Kokkinakis_2020_PoF,Guo_2022_PoF}.

\section{Conclusion}
This study introduces an in-house high-order compressible solver based on the compact finite-difference scheme, combined with high-order filtering and localized artificial diffusivity to enhance numerical robustness. The solver is subsequently verified and validated using a set of five canonical benchmark problems that span both discontinuous flows and turbulence-dominated regimes: the one-dimensional Sod shock tube, a two-dimensional shock–shear layer interaction, compressible turbulent channel flow, compressible turbulent boundary layer, and shock-turbulent boundary layer interaction over the compression ramp. The results demonstrate that the solver accurately captures the discontinuities without significant spurious oscillations, resolves shock-induced vortical structures, and reproduces key turbulent statistics, including mean profiles and second-order moments, in close agreement with exact solutions, experiments, and reference DNS/LES data. Overall, the results support the accuracy and robustness of the present high-order solver, providing a reproducible baseline for future developments and for high-fidelity simulations of compressible turbulent flows with shock interactions.

\section*{Acknowledgement}
This study is supported by National Supercomputing Center with supercomputing resources, including technical support (KSC-2023-CRE-0052).

\bibliographystyle{unsrt}
\bibliography{ref}

\end{document}